# WINDOWS KERNEL HIJACKING IS NOT AN OPTION: MEMORYRANGER COMES TO THE RESCUE AGAIN

Igor Korkin, PhD
Independent Researcher
Moscow, Russian Federation
igor.korkin@gmail.com

## ABSTRACT

The security of a computer system depends on OS kernel protection. It is crucial to reveal and inspect new attacks on kernel data, as these are used by hackers. The purpose of this paper is to continue research into attacks on dynamically allocated data in the Windows OS kernel and demonstrate the capacity of MemoryRanger to prevent these attacks. This paper discusses three new hijacking attacks on kernel data, which are based on bypassing OS security mechanisms. The first two hijacking attacks result in illegal access to files open in exclusive access. The third attack escalates process privileges, without applying token swapping. Although Windows security experts have issued new protection features, access attempts to the dynamically allocated data in the kernel are not fully controlled. MemoryRanger hypervisor is designed to fill this security gap. The updated MemoryRanger prevents these new attacks as well as supporting the Windows 10 1903 x64.

## 1. INTRODUCTION

The security of users' data and applications depends on the security of the OS kernel code and data. Modern operating systems include millions of lines of code, which makes it impossible to reveal and remediate all vulnerabilities. Attackers can exploit the OS vulnerabilities to perform malicious actions. Windows OS kernel remains one of the most desired targets for hackers.

Another big challenge of OS kernel protection is the third-party kernel-mode drivers, which execute at the same high privilege level as the OS kernel, and they also include a variety of vulnerabilities. Researchers consider that "kernel modules (drivers) introduce additional attack surface, as they have full access to the kernel's address space" (Yitbarek and Austin, 2019).

At the recent DEF CON hacking conference researchers from Eclypsium released a list of more than 40 drivers from Microsoft-certified hardware vendors, which are prone to privilege escalation attacks (Jesse and Shkatov, 2019).

Another vulnerability in a signed third-party driver was presented at the Blue Hat IL conference by security experts from the Microsoft Defender ATP Research Team. The vulnerable driver uses a watchdog mechanism based on user APC injection, which can also be exploited by attackers to bypass driver signature enforcement and gain escalated privileges. (Rapaport, 2019).

Recently revealed Banking trojan "Banload", which targets bank customers in Brazil and Thailand, applied a malicious kernel-mode component to fight with anti-malware and banking protection programs. This digitally signed malware driver is designed "to remove software drivers and executables belonging to anti-malware and banking protection programs", such as AVG, Avast, IBM Trusteer Rapport (Bisson, 2019; Kremez, 2019).

Kernel-mode drivers were also used during the recent RobbinHood ransomware attack. Hackers installed a legitimate driver and exploited its vulnerability to temporarily disable the Windows OS driver signature enforcement. Finally, they installed a malicious kernel driver (Cimpanu, 2020).

Notorious cryptocurrency mining malware also applies kernel-mode rootkits to prevent them from being terminated. Windows-based crypto miner infected more than 50000 servers from 90 countries (Harpaz and Goldberg, 2019; O'Donnell, 2019).

The Microsoft Security team do their best to maintain a high level of OS kernel protection by issuing various security features, for example, Microsoft Kernel Patch Protection (KPP) aka PatchGuard etc. At the same, time security researchers and rootkit developers are discovering different techniques to bypass PatchGuard. The most notable of them was GhostHook, which abused the Intel Processor Trace (PT) feature to overcome PatchGuard and patch the kernel. Cimpanu (2019) underlines that two recently published bypassing techniques InfinityHook and ByePg "establish a permanent foothold in the kernel itself and open the door for the return of rootkits on Windows 10".

We can see that on the one hand, all drivers and the OS kernel share the same memory space, and on the other hand, there are no built-in mechanisms to restrict access to the kernel memory. All drivers have full access to the system and can be used by attackers. Windows security features provide limited kernel memory protection.

**Threat model**

Let us assume that using various approaches, intruders are able to execute malicious kernel code. This paper analyses two types of attacks on kernel data, which result in the following, see Figure 1:

- gaining access to the files open in an exclusive mode (Handle Hijacking and Hijacking NTFS data structures);

- escalating process privileges without using the token swapping technique (Token Hijacking).

For the attacks on files, a legal driver creates a file via ZwCreateFile with zero flag ShareAccess, which gives the caller exclusive access to the open file. While the file remains opened all attempts to gain access to this file via ZwCreateFile are in vain. Windows OS detects this illegal access and returns a status sharing violation code (0xC0000043), which indicates that "a file cannot be opened because the share access flags are incompatible" (Microsoft, 2019).

This research reveals two different attacks, which bypass Windows security features and successfully gain unauthorized access to the files opened without shared access by patching OS internal data structures, related to the Object Manager and NTFS driver components.

The third attack escalates process privileges by patching the static and the variable portions of _TOKEN structure, without using token swapping or token stealing techniques. This type of attack is mapped to MITRE ATT&CK (2020) under Access Token Manipulation. The type of escalation privilege attack based on SeImpersonatePrivilege function is out of the scope of this paper (Bisht, 2020).

All newly proposed attacks are working transparently on Windows 10 1903 64 bit as well as for its security features, such as Patch Guard, Device Guard, and Security Reference Monitor.

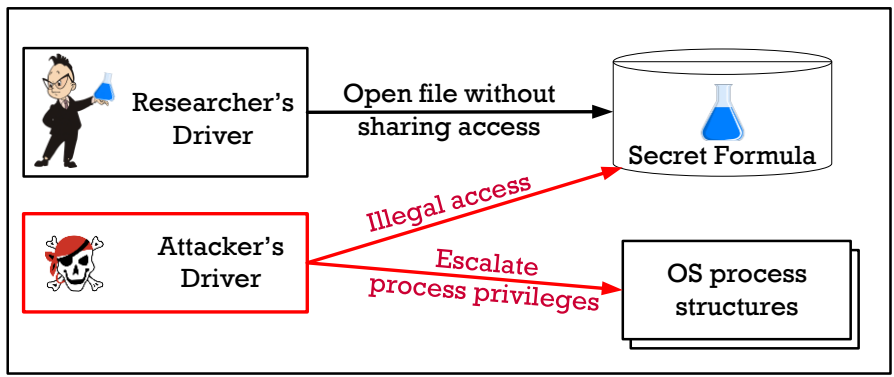

Figure 1. The following attacks will be considered: attacks on files and a privilege escalation attack.

To prevent all these attacks on Windows OS kernel data the updated MemoryRanger hypervisor will be presented.

MemoryRanger prevents attacks on files by running newly loaded drivers in separated kernel spaces as well as restricting access to the corresponding sensitive memory areas.

The newest key feature of MemoryRanger allows it to run a special data enclave for sensitive OS kernel data, such as _TOKEN structures. This enclave includes these sensitive OS structures, OS kernel core, and a limited number of OS kernel built-in drivers. This new scheme prevents illegal access from all drivers whether loaded before and after MemoryRanger.

The remainder of the paper is as follows.

Section 2 provides the details of the control flow and corresponding internal data structures involved during file operations in kernel mode. Two examples of hijacking attacks on files will be given.

Section 3 presents the details of access control issues in the Windows OS kernel and shows how attackers can hijack the corresponding structure in order to escalate the process privileges.

Section 4 contains the details of adapting MemoryRanger to prevent these attacks.

Section 5 and Section 6 focus on the main conclusions and further research directions respectively.

## 2. TWO HIJACKING ATTACKS ON THE FILES OPENED WITHOUT SHARED ACCESS

This section describes the internals of file operations in the Windows OS kernel: data structures and correlations between them. Two given hijacking attacks make it possible to illegally read and overwrite the content of the file opened in an exclusive mode. These two hijacking attacks are based on modifying the OS internal data structures involved in file operations.

### 2.1. Control Flow and Internal Data Structures Involved in Read and Write File Operations

Windows OS provides four main kernel API routines to create, read/write, and close files: ZwCreateFile, ZwReadFile, ZwWriteFile, ZwClose.

During file operations, several OS kernel components are involved (Russinovich, Solomon, and Ionescu, 2012; Tanenbaum and Bos, 2014). Each time a driver calls ZwCreateFile the control goes to the following OS kernel subsystems: I/O manager, Object Manager, Security Reference Monitor, NTFS driver, and finally, the control goes to the low-level drivers, such as Disk Filter Driver and Disk Class Driver. These are in charge of access to the physical disk.

The key features revealed by Korkin (2019) are that Security Reference Monitor checks access rights to the file for ZwCreateFile routine, while routines ZwReadFile, ZwWriteFile are uncontrolled by the Security Reference Monitor.

Once a file is created via calling ZwCreateFile, the OS creates a file handle, adds an entry to the Handle Table, allocates file object, NTFS data, and other structures. The created file handle is returned to the caller and is used as a key to read and write the open file using functions ZwReadFile and ZwWriteFile.

The details of the control flow and internal data structures involved in read and write file operations are given in Figure 2. Using a file handle, the OS traverses through the handle table to acquire the file object. By reading file object fields the OS locates control block structures (NTFS data structures) and moves to them. Disk drivers access the opened file on a disk by using NTFS data structures.

OS kernel treats read and write file access by traversing through these structures without any checks by Security Reference Monitor.

This vulnerability can be used to gain full illegal access to the files opened without shared access. To achieve it, intruders can create a file hijacker and patch any of the structures, see Figure 2.

As a result, all intruders' access attempts using the file hijacker handle will be redirected by the OS to the secret file, see Figure 3. This is the key point of all hijacking attacks on files.

Intruders can modify the following data to change the control flow, the corresponding attacks are in brackets:

- handle table entries (Handle Table Hijacking);
- file object (Hijacking FILE_OBJECT);
- NTFS data structures (Hijacking NTFS structures).

A File Object Hijacking attack was presented by Korkin (2019).

The next two subsections will describe the details of Handle Hijacking and Hijacking NTFS structures.

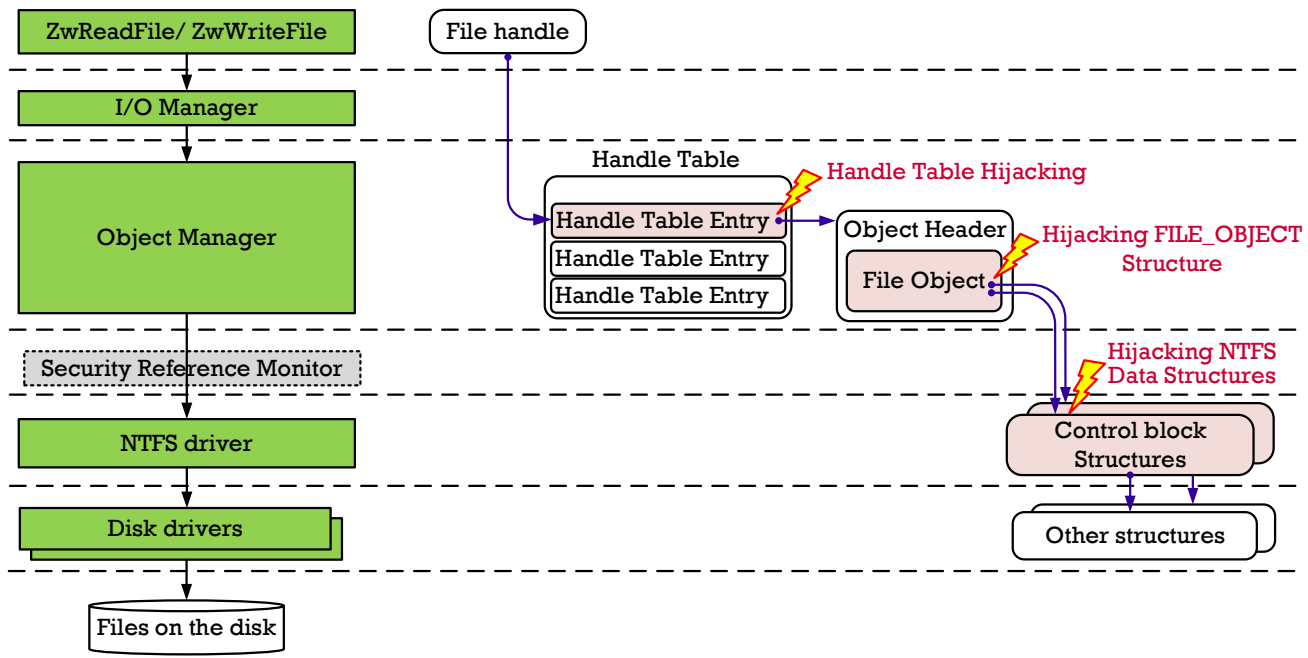

Figure 2. OS subsystems and corresponding data structures involved during read and write file operations.
(Russinovich, Solomon, and Ionescu, (2012), part2, pp 441).

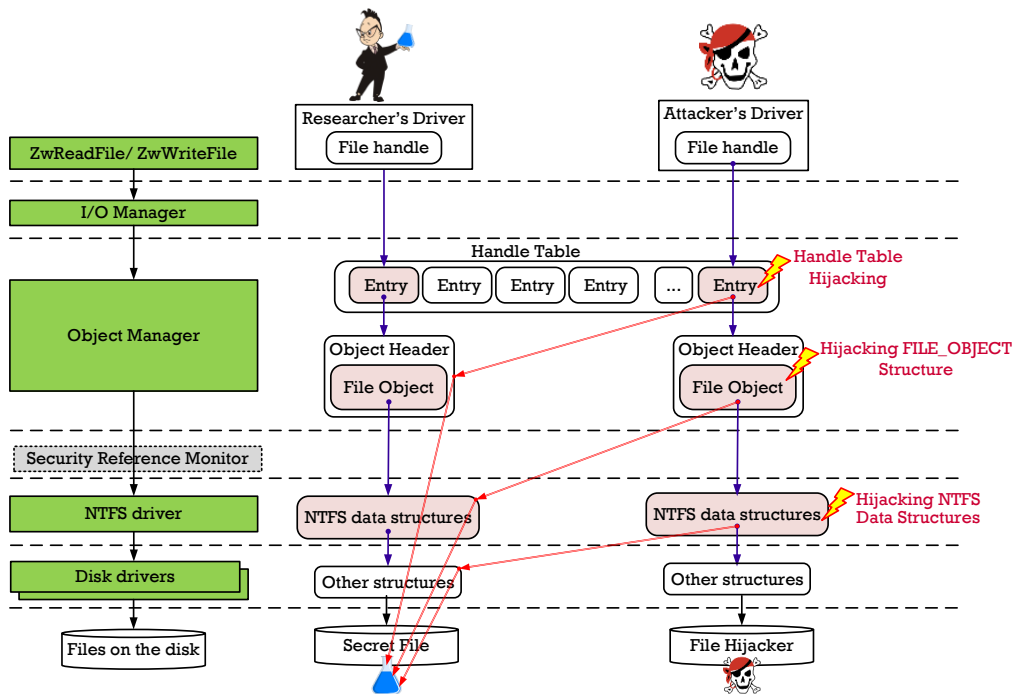

Figure 3. An attacker creates a file hijacker and applies three different hijacking attacks: Handle Table Hijacking, Hijacking FILE_OBJECT structure, Hijacking NTFS structures, which are based on patching handle table entries, file object, and NTFS data structures.

### 2.2. Handle Table Hijacking

This section describes the details of the Kernel Handle Table and how attackers can hijack its values to gain illegal access to the files.

### Handle Table Basics

The security issue with access to the files, locked by an application from another application is typical for Windows OS. Sysnap (2011) describes the details of illegal access to the locked file by modifying the handle table, belonging to the application. The author's approach is designed for the user-mode handles and process for Windows XP and 2003. This section describes how to adapt Sysnap's idea of patching the handle table for kernel case in the most recent Windows OS.

A way for patching handle table entries for user process in order to change handle access rights is implemented in Blackbone by DarthTon (2019-a).

The Kernel Handle Table is used by the Windows OS to store the mapping from the handles to the corresponding object structures (Tanenbaum and Bos, 2014; Probert, 2010; Schreiber, 2000). Using a handle, the OS traverses through Kernel Handle Table to acquire the object. Exported symbol nt!ObpKernelHandleTable points to this table. The address of this table can also be gained by reading the field EPROCESS.ObjectTable for SYSTEM:4 process. Kernel Handle Table is involved each time a driver reads and writes a file. This handle-based mechanism manipulates various objects, such as files, processes, threads, or registry keys.

The presented research is focused only on handles related to the file system, but the achieved results can be applied to all kernel objects as well.

For each newly created file, the kernel handle table has an entry and an index of each entry equals the returned handle value (Hale-Ligh, M. Case, A, Levy, J., Walters, 2014). Each entry is defined in a HANDLE_TABLE_ENTRY structure, which includes access rights granted to the object (field GrantedAccessBits) and the link to the created object (field ObjectPointerBits). The field ObjectPointerBits includes 44 low bits of OBJECT_HEADER address, which can be used to gain the FILE_OBJECT address (Monnappa, 2018; CodeMachine, 2019).

Tanenbaum and Bos (2014) found that "system calls, like ZwReadFile and ZwWriteFile, use the kernel handle table created by the object manager to translate a handle into a referenced pointer on the underlying object, such as a file object, which contains the data that is needed to implement the system calls" (pp 899).

Handle table can have several levels, the number of levels and the number of entries in each level depends on which Windows version is being used (Suma, Dija, Thomas, 2014; Probert, 2010; Schreiber, 2000). Windows OS provides a function ExEnumHandleTable to enumerate all the valid handles in a handle table. ExEnumHandleTable specifies an enumeration callback function, which is called for each valid handle in the handle table (DarthTon, 2019-b; Treadwell, 1989). The enumeration procedure needs to release implicit locks for each handle via call ExUnlockHandleTableEntry (ReactOS, n.d.; WRK. n.d.). The enumeration procedure returns a bool value. To stop the enumeration the procedure needs to return a TRUE value and as a result ExEnumHandleTable also returns TRUE. To continue the enumeration, the procedure needs to return FALSE.

Using ExEnumHandleTable intruders can access the handle table entry, which belongs to the file hijacker, and patch it.

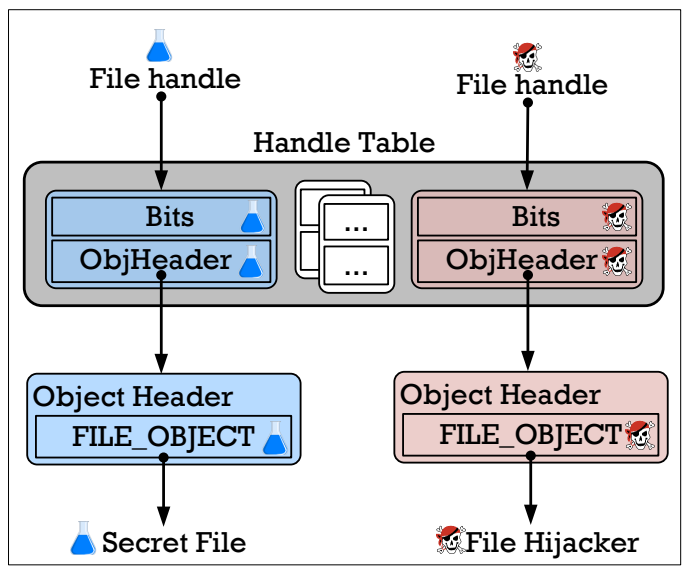

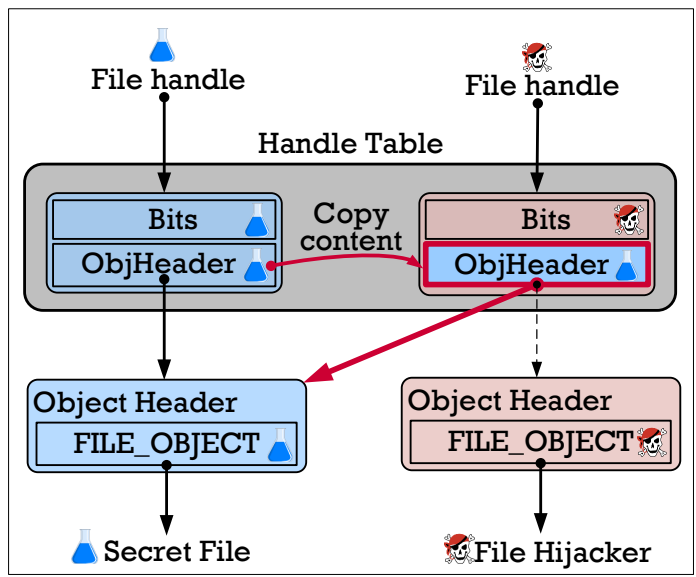

Figure 4. The control flow between files structures: *a)* before and *b)* after Handle Table Hijacking

**The Algorithm of Handle Table Hijacking**

The research reveals that during read and write access OS traverses through the kernel handle table and acquires the file object, without any checks. Attackers can use this vulnerability to gain illegal access to the exclusively open file in this way:

1. Reveal OBJECT_HEADER address of the secret file.
2. Create a file hijacker and locate a corresponding entry in the handle table entry.
3. Overwrite the ObjectPointerBits field in this entry using the OBJECT_HEADER address of the secret file.

After this handle hijacking attack all read and write access using a hijacked file handle will be redirected to the secret file, see Figure 4.

This hijacking attack requires overwriting just 44 bits of dynamically allocated data, which is enough to gain illegal access to the exclusively opened file. This redirection will be carried out by the Windows OS automatically and transparently for the built-in security features.

On the one hand, intruders can tamper with entries of the kernel handle table in order to exploit the translation mechanism, and on the other hand, Windows security features do not check the integrity of this table and cannot reveal this attack.

## 2.3. Hijacking NTFS data structures

This section describes the details of the attack called "Hijacking NTFS data structures".

This attack is an improvement of the attack presented by Korkin (2019), which was based on Hijacking FILE_OBJECT. Let us assume, that a security service continuously provides integrity and confidentiality for all FILE_OBJECT structures. As a result, only OS kernel has access to the FILE_OBJECT structures, while access attempts from all other drivers are forbidden.

In this new situation, attackers cannot use FILE_OBJECT hijacking attack and they need to prepare a new attack. Attackers decide to organize a new lower-level attack on control block structures or NTFS data structures, which FILE_OBJECT fields point to, see Figure 4 *b)*.

FILE_OBJECT structure includes fields FsContext and FsContext2, which point to the control block structures: FsContext member points to the File Control Block, FCB (Stream Control Block, SCB) and FsContext2 points to the Context Control Block, CCB. Each file stream is uniquely represented in memory by an FCB structure. CCB structure is created by file system drivers to represent an open instance of a file stream (Nagar, 1997). This mechanism is deeply integrated into the Windows OS kernel and is very rarely updated.

FsContext and FsContext2 represent the physical stream context and the user handle stream context. FsContext2, is used to point to the Channel Control Block or CCB (Miller, 1991; Probert, 2004).

Let us move on to the details of FCB and CCB structures. These structures pointed by FsContext and FsContext2 are only partially documented, but at the same time, the research has revealed the following details.

The definition of the SCB, FCB and CCB for Windows NT 4.0 are in the file "ntfsstru.h" (Microsoft, n.d.-a). The definition of these structures can also be found in file "cdstruc.h" from ReactOS (ReactOS, n.d.-a). These definitions can be used for understanding some basic file principles because the structures are partially updated in the most recent Windows.

According to the MSDN file object's FsContext member stores a pointer to the FSRTL_ADVANCED_FCB_HEADER structure, which uniquely identifies the file stream to the file system (MSDN, 2018-a, MSDN, 2018-b, MSDN, 2018-c).

The research shows that an appropriate target for this new attack is a FSRTL_ADVANCED_FCB_HEADER structure, which FsContext fields from a FILE_OBJECT structure points to.

In addition, the research has revealed that fields FsContext and FsContext2 point to the contiguous memory blocks and, when using a hijacking attack, intruders can copy and overwrite the content of these two structures simultaneously.

Structures FSRTL_ADVANCED_FCB_HEADER are not protected by the OS security features, and their patching does not cause any kernel security check failure errors, such as BSOD.

In a nutshell, Hijacking NTFS data structures is based on locating internal file object structures from FsContent and FsContent2 fields, copying their content to the corresponding memory areas pointed by the fields of file object hijacker and additional patching. Without this patching during read or write operation, the OS detects aforementioned copying and causes BSOD with RESOURCE_NOT_OWNED (0x000000E3) bug check.

**The Algorithm of Hijacking NTFS data**

To implement Hijacking NTFS data structures, intruders have to locate the NTFS data structures, which correspond to the secret file and to the file hijacker, and engage in the following three steps:

**Step 1.** Overwrite the content of attackers' FSRTL_ADVANCED_FCB_HEADE

R structure using the data from the FSRTL_ADVANCED_FCB_HEADER structure, which belongs to the secret file.

However, this overwriting is not enough, because Windows OS reveals that a malware driver's thread tries to release a resource it did not own and Windows OS causes BSOD with RESOURCE_NOT_OWNED bug check. To overcome this BSOD attackers move to the second step.

**Step 2.** Set attackers' thread ID gained by PsGetCurrentThread to the following fields in FSRTL_ADVANCED_FCB_HEADER structure:

- Resource->OwnerEntry.OwnerThread;
- PagingIoResource->OwnerEntry.OwnerThread.

This patching helps malware driver to overcome BSOD with RESOURCE_NOT_OWNED bug check.

Windows OS kernel changes the content of FSRTL_ADVANCED_FCB_HEADER structure while returning the result of reading and writing to the driver and if attackers try to access the file using previously modified structure the Windows OS will cause BSOD again. If attackers want to access the secret file several times, they move on to Step 3.

**Step 3.** Repeat Step 1 and Step 2 before each read and write access attempt during every hijacking attack.

Attackers have to repeat Step 1 and Step 2 before each unauthorized read and write access attempt, thus preventing the aforementioned BSOD.

As a result, each time attackers read and write a file using a hijacked file handle, OS walks through patching structure and provides illegal read and write access to the secret file, without BSOD.

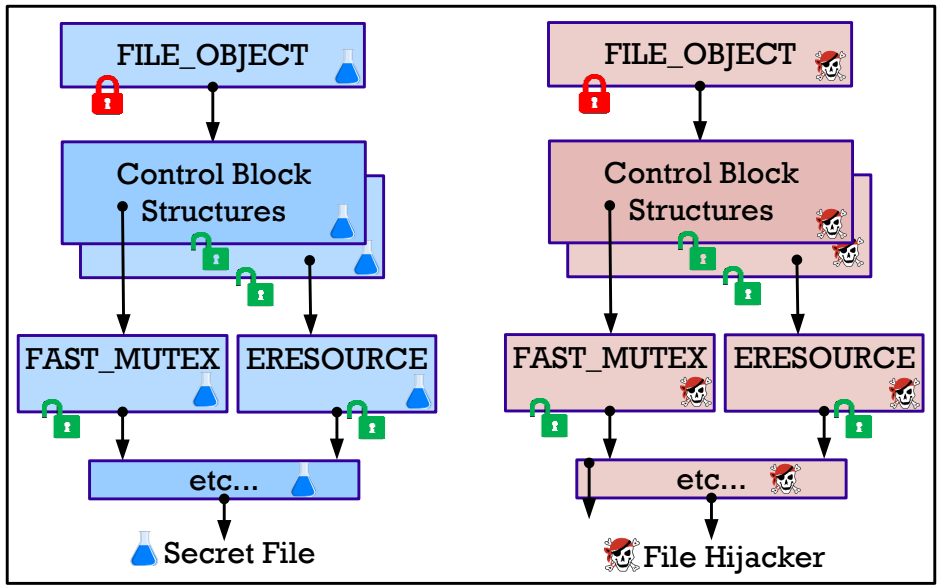

*a)*

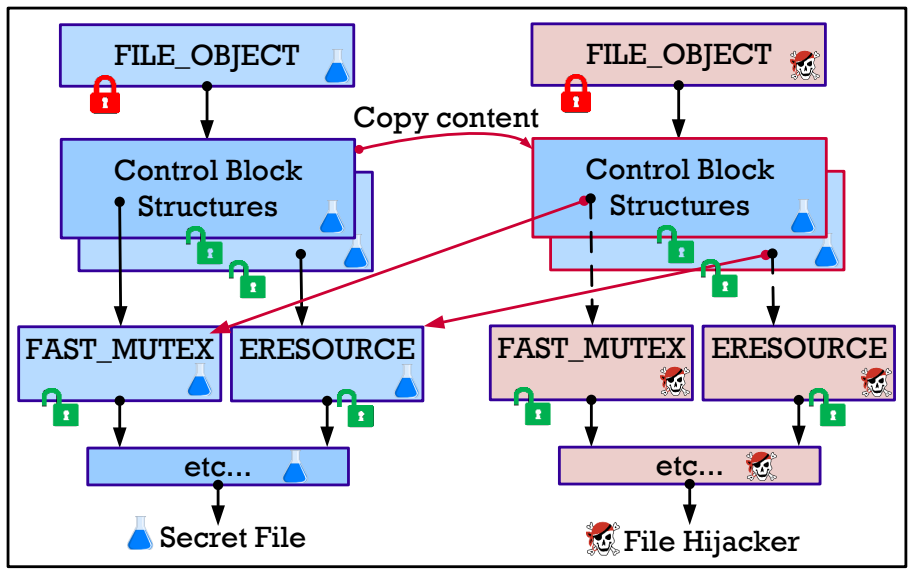

*b)*

Figure 5. The control flow between files structures: a) before and b) after hijacking the NTFS control block structures, e.g. FSRTL_ADVANCED_FCB_HEADER

## 3. TOKEN HIJACKING ATTACK: WHAT AND HOW

The process privilege mechanism is crucial for OS security. This section describes the process privileges mechanism and how it can be hijacked by modifying the content of dynamically allocated memory.

For each newly created process, Windows OS allocates a new EPROCESS structure and adds it to the list. This structure includes internal information about this process: its name and ID, threads and handles details, etc. (Monnappa, 2018; Tanenbaum and Bos, 2014). _TOKEN structure describes the process access token, which contains the security-related information about the process: user's and group SIDs, process privileges, etc. (Hoglund and Butler, 2006). The _TOKEN structure is pointed by the field Object, which is located in Token _EX_FAST_REF in _EPROCESS structure, see Figure 6.

Windows OS provides discretionary access control, which is governed by two main parts (Ismail, Aboelseoud, and Senousy, 2014; Johnson, 2015):

- an access token associated with each process;
- a security descriptor associated with each object, such as a file.

According to the Russinovich (1998) and Stallings (2002) each time a process tries to access the object; the Security Reference Monitor reads the SIDs and group SIDs from _TOKEN structure to determine whether or not this access is allowed.

Attackers apply various techniques to elevate privileges for the malware process (Chebbi, (2019).

API-based approach to steal tokens in Windows was proposed by Barta (2009). The author leverages several kernel API routines, which makes it difficult to apply this approach via the malware payload.

The ideas of applying direct kernel object manipulation (DKOM) to the process token in order to gain elevated access were discussed by Hoglund and Butler (2006) more than 10 years ago.

On the Black Hat USA 2004 for the first time, they proposed an idea of adding groups to Token structure using DKOM (Hoglund and Butler, 2004). The authors' idea is based on patching UserAndGroups array so that the required high privileges will be enabled for the process.

To prevent these manipulations Windows experts moved one step ahead and since Windows kernel 6.x several fields such as SidHash and RestrictedSidHash have been added into the _TOKEN structure to provide the integrity of this structure. OS checks these hashes to ensure that the SID list is not patched. These new fields prevent attackers from directly modifying the SID list.

Perla and Oldani (2010, pp. 295) underlined three alternatives to bypass this security hash-based barricade. One of them is *token stealing or token swapping* and it is based on overwriting the Object field in the _EPROCESS structure from the malware process. This uses the value from the _EPROCESS structure corresponding to the higher-privileged process, for example, SYSTEM:4 (Perla and Oldani, 2010, pp 305). However, newly updated Microsoft Windows Defender Antivirus detects such escalation by monitoring token-swapping attempts (Oh, 2017; Singh, Kaplan, Feng, and Sanossian, 2019). Bui (2019) shows that access token manipulation can be detected using auditpol, which is based on ETW, but this detection approach can also be tampered with due to attacks on ETW.

**A New Token Hijacking Attack**

I propose a new Token Hijacking Attack, which is a development of ideas of Hoglund and Butler (2006). In a nutshell, attackers need to escalate privileges so that the calculated SidHash value will be corrected and the integrity check will not reveal any changes.

Attackers can achieve this by overwriting the following whole three fields using the corresponding values from the Token structure corresponding to the higher-privileged process:

- UserAndGroupCount;
- UserAndGroups array: Attributes and Sid structures;
- SidHash structure;

*The key feature is to completely copy the UserAndGroups array with updated internal structure arrangement* from the Token for higher privilege process, while Hoglund and Butler (2006) proposed to overwrite just a few fields.

During this attack copying UserAndGroupCount field and SidHash structure is trivial because they have the same size while copying a variable part pointed by UserAndGroups is quite complicated. The number of entries in UserAndGroups and sizes of SID structures are not the same for various processes with different credentials, Figure 6.

The following two facts make this attack possible. Firstly, this updating is more than enough to gain elevated privileges yet not being detected by the OS. Secondly, TOKEN structure for common processes always has enough space, because a variable portion of _TOKEN structure for System:4 process is less than the corresponding structure for a common one.

This attack has been successfully tested on the newest Windows 10 1903 x64, the source code and demo in this paper (Korkin, 2020).

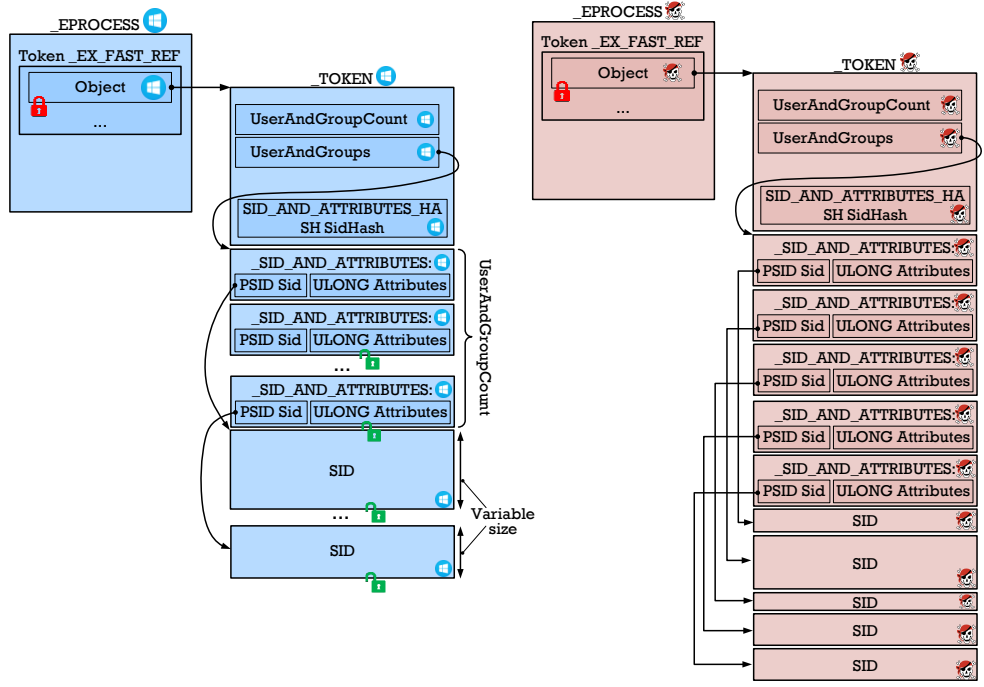

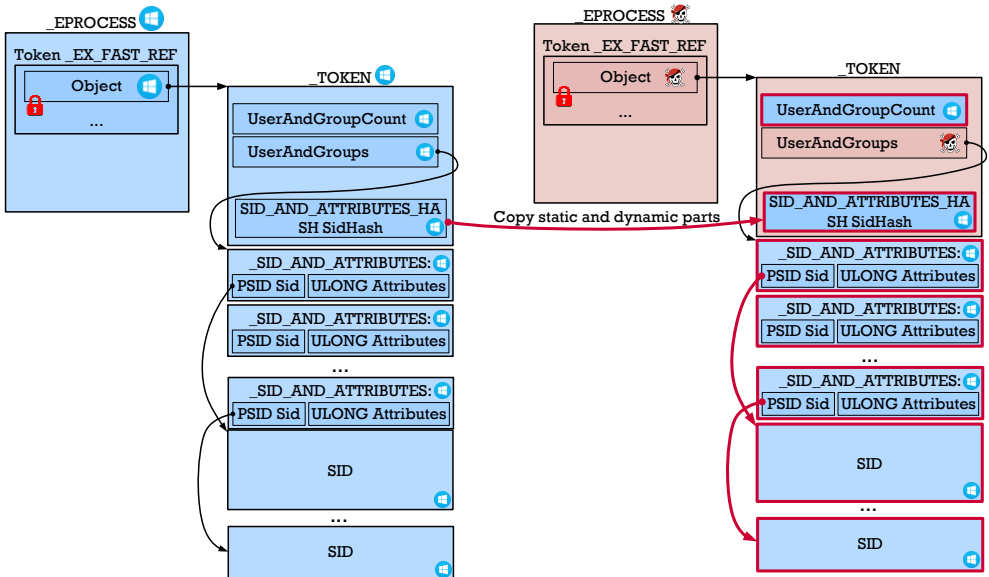

Figure 6. The content of _EPROCESS and _TOKEN structures for SYSTEM:4 and malware processes: *a)* before and *b)* after Token Hijacking

# 4. MEMORYRANGER PREVENTS KERNEL HIJACKING STRUCTURES

This section describes the details of how the updated MemoryRanger hypervisor prevents hijacking attacks (Korkin, 2018; Korkin, 2019).

## 4.1. MemoryRanger Overview

MemoryRanger is a hypervisor-based solution (a bare-metal hypervisor), designed to provide integrity and confidentiality for kernel-mode code and data. MemoryRanger leverages Intel VT-x technology and Extended Page Tables (EPT).

MemoryRanger protects kernel memory by using isolated kernel enclaves with specified memory access restrictions.

By running kernel drivers in separate memory enclaves MemoryRanger protects kernel memory from being tampered with:

- It prevents attacks on OS kernel code and data from newly loaded drivers;
- It protects the code and data of newly loaded drivers from the attacks from each other.

After loading, MemoryRanger allocates the default kernel enclave: OS kernel and all drivers loaded before are running inside this enclave.

Newly loaded drivers are running in separate enclaves. MemoryRanger traps the loading of each new driver and allocates an isolated kernel enclave for this driver. Each newly loaded kernel driver is running only inside the corresponding allocated kernel memory enclave.

MemoryRanger updates the memory access restrictions for each enclave in run-time, which makes it possible to protect sensitive memory areas. MemoryRanger can monitor access to the kernel-mode memory and redirect the illegal access to the fake page.

## 4.2. MemoryRanger: Key components

MemoryRanger has the following key components:

- A kernel-mode driver;
- DdiMon;
- MemoryMonRWX;
- Memory Access Policy (MAP).

All the details about MemoryRanger components are given by Korkin (2019).

MemoryRanger does the following:

- registers a driver-supplied callback that is notified whenever a new process is created/deleted and an image is loaded;
- hooks kernel API by using DdiMon component;

- restricts memory access even to a byte by using MemoryMonRWX;
- provides dynamically updated access control rules using MAP.

### 4.3. Main Updates of MemoryRanger to Block New Hijacking Attacks

In order to prevent the newly presented hijacking attacks, the following modifications have been added to the MemoryRanger.

MemoryRanger prevents Handle Hijacking and Hijacking NTFS data structures by doing the following:

- it hooks ZwCreateFile() and ZwClose routines to locate the involved data structures in memory;
- To prevent Handle Hijacking, it locates the HANDLE_TABLE_ENTRY structure corresponding to the opened file using the ExEnumHandleTable routine;
- To prevent Hijacking NTFS Data, it locates the FSRTL_ADVANCED_FCB_ HEADER structure using the pointers from FILE_OBJECT.
- MemoryRanger restricts access to the structures in the corresponding enclaves.

To prevent Token Hijacking MemoryRanger implements a new technique, which is based on allocating a *special isolated enclave, which includes only sensitive kernel data*, see Figure 7.

The details of these prevention techniques are given below. The source code of updated MemoryRanger, attacker, and allocator drivers as well as video demos are here (Korkin, 2020).

### 4.4. Details of Prevention of Handle Hijacking Attack

Preventing Handle Hijacking requires a fine-grained approach.

MemoryRanger prevents Handle Hijacking by blocking only write access to ObjHeader field, which has 6 bytes and corresponds to the file object header, see Figure 7. Neither does MemoryRanger restrict read access for ObjHeader, nor does it prevent any access to other fields of this entry, because they are used by the OS. In fact, some fields of these entries have to be accessed for write attempts due to synchronization issues and their restriction causes BSOD.

MemoryRanger is notified about creating a new file by hooking ZwCreateFile routine and next it locates handle table entry by using file handle and the ExEnumHandleTable routine.

### 4.5. Details of Prevention of Hijacking NTFS data structures

For Hijacking NTFS data structures intruders modify the control block structures (FSRTL_ADVANCED_FCB_HEA

DER), which correspond to the file hijacker.

To prevent this attack MemoryRanger implements a similar approach based on locating control block structures and restricting access to them.

### 4.6. Details of Prevention of Token Hijacking

Token Hijacking Attack is tampering with static and dynamic parts of _TOKEN structures, which results in local privilege escalation.

To block Token Hijacking a special isolated kernel enclave is allocated to host sensitive data. This new enclave includes only sensitive kernel data, such as _TOKEN structures, Windows kernel core (ntoskrnl.exe), and a limited number of trusted Windows drivers. All other drivers will be excluded from this enclave, see Figure 7.

This new scheme isolates token structures from all drivers loaded after and even before MemoryRanger without restricting the OS kernel.

MemoryRanger is notified about creating a new process by registering a callback routine via call PsSetCreateProcessNotifyRoutineEx.

### 4.7. Empirical Test Results

All these attacks and their prevention have been successfully tested on Windows 10 1903 x64, details are in (Korkin, 2020).

### 4.8. Performance Impact

MemoryRanger causes affordable performance degradation. Switching between kernel enclaves is the main problem of this performance degradation. Changing the EPT pointer causes the flushing of TLB and further filling the TLB. Details about measuring the performance were given previously by Korkin (2018). I can conclude that MemoryRanger is suitable to protect the rarely accessed safe areas. To avoid this degradation, the new version will support VPID, which is designed to meet this need.

### 4.9. MemoryRanger vs. Virtual Secure Mode

One of the global security challenges for modern operating systems is to prevent illegal access to the kernel data from drivers, while all drivers and OS share the same memory space.

MemoryRanger is designed to tackle this issue by isolating newly loaded drivers inside allocated separated memory enclaves from the rest of the OS kernel. This drivers' isolation can prevent attacks from kernel rootkits as well as providing exploit mitigation.

MemoryRanger can be applied to protect Unix-based systems running on AMD and ARM CPUs.

MemoryRanger includes a kernel driver, which allows it to trap and parse OS-related events. Using a hypervisor component

MemoryRanger restricts access to the memory transparently for the OS kernel. MemoryRanger is protected from kernel attacks, due to running in ring -1)

Windows OS comprises a new technology called Virtual Secure Mode (VSM), which is designed to maintain a secure Windows environment. *VSM provides a particular case of enclave-based protection with only two memory partitions* called VTL0 and VTL1, while *MemoryRanger implements a general case with an infinite number of kernel enclaves*. MemoryRanger has been tested using three (Korkin, 2019), four (Figure 7), and *five separate enclaves* (Korkin, 2018).

## 5. CONCLUSION

To sum up I would like to highlight the following:

1. Windows OS kernel manipulates dynamically allocated data, which can be tampered with by intruders during cyberattacks. Windows security features provide integrity only for limited memory areas, while others are becoming susceptible.
2. Two new presented attacks on files: Handle Hijacking and Hijacking NTFS structures make it possible to gain illegal access to the files opened in an exclusive mode bypassing Security Reference Monitor.
3. Hijacking Attack on NTFS data structures has never been presented before.
4. A new Token Hijacking attack results in process privilege escalation via copying SID with their attributes as well as SID hashes from a higher privileges process. This attack gains elevated privileges with the correct hash value.
5. Updated MemoryRanger prevents attacks on files by running drivers inside isolated enclaves and restricting access to the corresponding data structures.
6. To prevent Token Hijacking, updated MemoryRanger implements a new special enclave, which includes only sensitive data and a part of the Windows OS kernel; all other drivers as well as newly loaded ones are not able to tamper with this data.
7. All new attacks on files and tokens have been successfully tested on the most recent Windows 10 1903. Updated MemoryRanger can prevent all mentioned hijacking attacks.
8. Various cybersecurity solutions will benefit from applying MemoryRanger. The source code and all demos are uploaded here (Korkin, 2020).

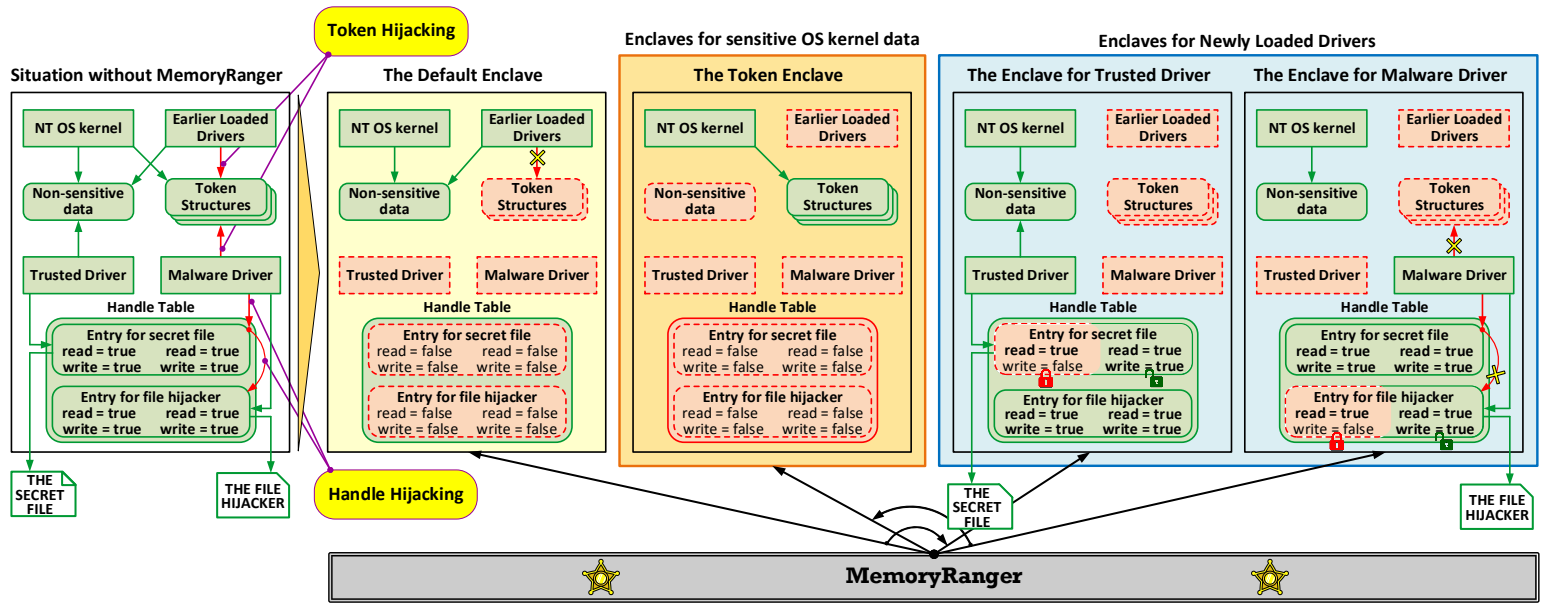

Figure 7. MemoryRanger prevents Handle Hijacking and Token Hijacking

# 6. FUTURE PLANS

Updated MemoryRanger is a very promising project and the following future directions can be outlined.

## 6.1. Prevent patching of OS Internal Structures by Intel Memory Protection Keys (MPK)

Prakash, Venkataramani, Yin, and Lin (2013) revealed the following examples of how attackers can patch the OS internal structures to gain ongoing and undetectable access to the target system:

- Hackers can prevent a malicious process from terminating by changing the corresponding EPROCESS.Flags to 0xFFFFFFFF value;
- Hackers can make a file completely inaccessible by changing the file name field in FILE_OBJECT to an empty string;
- Hackers can hide a program from the process list in the Task Manager by changing the corresponding EPROCESS.UniqueProcessId to 0.

Despite the fact that the results were checked on Windows XP, the similar principles can be expanded on the most recent Windows 10 and for more OS internal structures.

A recently issued key-based permission control called Intel Memory Protection Keys (MPK) can be leveraged to maintain and enforce memory access permission. MPK has several key benefits over the page-table mechanism (Park, Lee et al. 2019).

## 6.2. MemoryRanger can do Deep Inspection of Memory Access

The current version of MemoryRanger checks access rights using only the address of the module, which implements this access. As a result, malware can gain illegal access to sensitive data by exploiting any vulnerable function in the trusted module.

The idea is to verify the function call stack additionally to check which sequence of functions leads to access this sensitive memory area.

This new check helps to inspect memory access a bit deeper and prevent the mentioned attack.

Finally, intruders have to exploit not only the function which accesses sensitive data but also the whole sequence of functions, which is very difficult and time-consuming.

## 6.3. MemoryRanger vs. Mimikatz

One more advanced direction is to analyze and prevent activities of Mimikatz, which is an effective post-exploitation tool (Delpy, 2020).

Mimikatz applies its driver to provide various commands to play with kernel memory:

- read and write kernel-mode memory from user-mode applications;
- disable Protected Process Light (PPL) mode;
- duplicate process token;
- set all privileges for a process.

### 6.4. MemoryRanger vs. Privilege Escalation by using buggy StopZilla driver

Security researchers have revealed that StopZilla AntiVirus Software includes a vulnerable driver, which can be used to elevate process privilege.

For this attack hackers need to have SeLoadDriverPrivilege to load a vulnerable driver and launch the exploit to patch TOKEN structure.

Ideas and details of such attacks were discussed by Pierini (2019) at Hack In Paris conference and by Cocomazzi and Pierini (2020) at the HITBSecConf2020.

Protection of the OS kernel from vulnerable signed third-party drivers is a serious security problem, because these drivers run in the kernel memory without any restrictions.

MemoryRanger can be updated to run StopZilla's driver in isolated enclaves and prevent the overwriting of the sensitive data, such as TOKEN structures. We cannot reveal all vulnerabilities, but we can protect the sensitive data, which is usually the top target of attacks by hackers.

### 6.5. Isolated Enclave for the OS Scheduler: Unikernels Based Protection

The current version of MemoryRanger has an issue with the protection of data from being tampered with by drivers loaded before it.

MemoryRanger protects rarely accessed data with acceptable performance degradation, while protection of frequently accessed data causes significant performance degradation.

To overcome this performance degradation and protect sensitive OS kernel data, such as EPROCESS structures from all drivers, MemoryRanger can allocate the following enclaves:

- **the Scheduler enclave** includes sensitive OS structures, such as EPROCESS, ETHREAD etc., which are frequently used by OS kernel scheduler. OS scheduler is located in this enclave. Likewise, this enclave contains file system drivers and their structures. As a result, this enclave includes a minimal list of drivers and their structures.
- **the Default enclave** contains all other drivers loaded before MemoryRanger. This enclave excludes all OS internal data structures from the Scheduler enclave, while drivers from the Default enclave are excluded from the Scheduler enclave.

- **the Data-Only enclave** includes only sensitive data structures, which are rarely accessed by the OS kernel drivers, such as Token structures.
- **a new enclave** is allocated for each newly loaded driver.

This scheme could help to isolate sensitive OS kernel data from being tampered by the drivers, loaded before the MemoryRanger.

Another solution to this challenge is to load MemoryRanger at boot time; it can be solved by using UEFI hypervisor that supports booting an operating system (Tanda, 2020).

### 6.6. MemoryRanger and Hyper-V

The current version of MemoryRanger does not support concurrent execution with Hyper-V, which is a Windows built-in hypervisor.

This issue can be solved by installing Windows OS in a virtual CPU without VT-x/EPT support and further enabling VT-x/EPT support to run MemoryRanger. Another way is to manually disable or uninstall Hyper-V to run MemoryRanger.

However, A. Eremeev (2020) implemented a bare-metal hypervisor with VT-x/EPT support, which works well with enabled Hyper-V. This experience can be used to expand features of MemoryRanger.